\documentclass[aps,prl,twocolumn]{revtex4}

\usepackage{epsfig}
\begin{document}

\title{Insulating behavior of $\lambda$-DNA on the micron scale}
\author{Y. Zhang$^1$, R. H. Austin$^1$, J. Kraeft$^2$, E. C. Cox$^2$, and N. P. Ong$^1$}

\affiliation{$^1$Department of Physics, and $^2$Department of Molecular Biology, Princeton University, Princeton, 
New Jersey 08544}

\date{\today}

\begin{abstract}
We have investigated the electrical conductivity of $\lambda$-DNA using DNA covalently bonded to Au electrodes.  
Thiol-modified dTTP was incorporated into the `sticky' ends of bacteriophage $\lambda$-DNA using DNA polymerase.  
Two-probe measurements on such molecules provide a hard lower bound for the resistivity $\rho>10^6\,\mathrm{\Omega 
cm}$ at bias potentials up to 20 volts, in conflict with recent claims of moderate to high conductivity.  By direct 
imaging, we show that the molecules are present after the measurements.  We stress the importance of eliminating 
salt residues in these measurements.
\end{abstract}

\maketitle
The question whether DNA is electrically conducting has generated broad interest.  The initial spurt of interest 
arose in photoexcitation experiments which were interpreted in terms of long-range electron transfer~\cite{barton}.  
In the past few years, there have been upwards of 20 papers reporting the results of more direct electrical 
measurements ranging from contactless meaurements at microwave frequencies to DC measurements.  A distressingly wide 
range of conductivity values -- from $\rho<10^{-4}\,\mathrm {\Omega cm}$ to $\rho>10^6\,\mathrm{\Omega cm}$ -- has 
been reported~\cite{fink,braun,dekker,spanish,DekkerAPL}.  Proximity-induced superconductivity in DNA has also been 
claimed~\cite{superconducting}.  Recently, local polarization measurements by `electrostatic force microscopy' have 
been used to show that $\lambda$-DNA is insulating~\cite{spanish2,Sohn}.  We note, however, that the 
force-microscopy experiments probe conductivity at relatively weak bias potentials.  

In many of the DC measurements, contact with the metal electrodes (usually Au) was achieved by laying down the 
molecules directly on the electrodes. Although expedient, this approach raises several concerns.  It is very 
difficult to prove that the DNA molecule is in direct physical contact with the electrodes.  Even if contact is 
attained, the weak physical adhesion between DNA and Au may produce an insulating contact and possibly account for 
the wide variation in reported resistivities~\cite{sciencecomment}.  A recent experiment on 
octanedithiol~\cite{science} has shown that deliberate chemical bonding between organic molecules and metal 
electrodes is a pre-requisite for achieving reproducible conductivity results.  Thus a better approach would be to 
achieve direct chemical binding between the open ends of $\lambda$-DNA and Au.  The bonds should be strong enough to 
withstand shear forces in a flow, and should survive the measurement process.  A second concern is the shunting 
effect of buffer residue.  Because of its finite conductance, the buffer salts which coat the electrodes and 
substrate produce a spurious conductance signal.  Hence adequate salt removal is important.  We report the results 
of experiments performed along these lines. Our results show that $\lambda$-DNA is a good insulator up to bias 
potentials of 20 volts. 

Chemical binding between organic molecules and Au is usually achieved by the Au-thiol (SH) chemical 
bond~\cite{mirkin}.  Commercially available oligonucleotides modified to incorporate the thiol group usually have 
carbon-chain spacers (C3 or C6) between the thiol group and DNA~\cite{braun,DekkerAPL,glen}, which may present 
barriers to electron transfer.  To avoid the spacer problem, we adopted an approach in which the DNA base-pair is 
bound {\it directly} to gold electrodes by a Au-thiol bond.  This approach should provide the most direct 
conductance channel between the gold electrode and the putative electronic ``$\pi$-way'' proposed for the DNA 
helix~\cite{history}. 

\begin{figure}[h] 
\centerline{\epsfig{figure=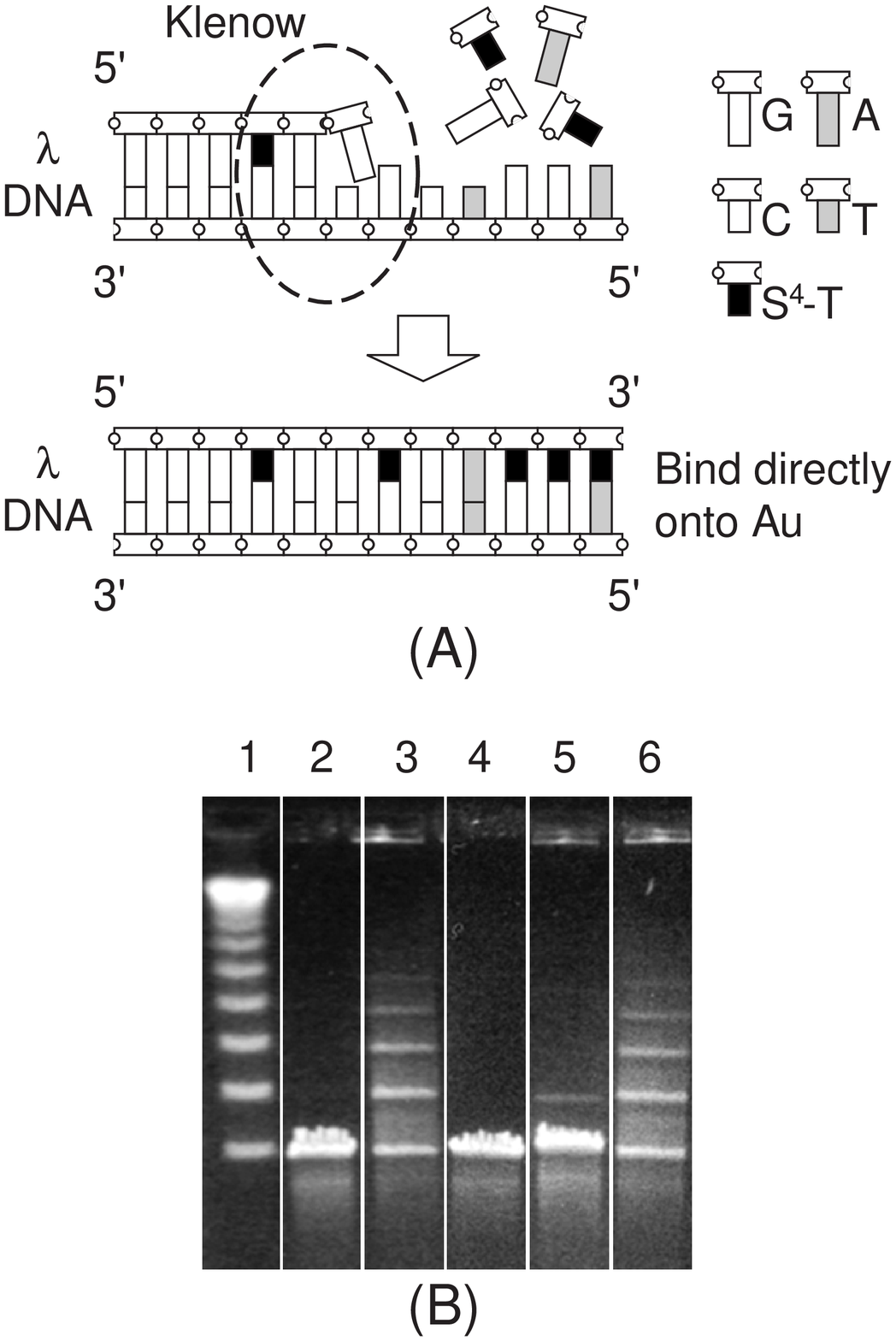,height=5in,width=3.3in,clip=0}}
\caption {\label{fig1} (A) The schematic of the incorporation of deoxynucleoside triphosphates into $\lambda$-DNA 
`sticky' ends using the Klenow fragment (3'$\rightarrow$5' exo$^-$); (B) Pulsed-field gel (PFG) electrophoresis of 
thiol-modified and natural $\lambda$-DNA before and after ligation~\cite{PFG}.  Lane 1: $\lambda$-DNA PFG marker 
(New England Biolabs); 2: unmodified ``natural-$\lambda$'' monomers; 3: ligated ``natural-$\lambda$'' multimers; 4: 
thiol-modified $\lambda$-DNA (``HS-$\lambda$'') monomers; 5: ``HS-$\lambda$'' after ligation, with no significant 
multimer formation; 6: ``control-$\lambda$'' after ligation, proving that the presence of unincorporated S$^4$-dTTP
and other dNTPs does not inhibit the ligation reaction.}
\end{figure}

$\lambda$-DNA is a double-stranded DNA helix comprised of 48,502 base pairs (length $\sim 16\,\mu$m).  At the 
extremities, there are single-stranded 12-base 5' overhangs (`sticky ends'), with the complementary sequences 
$$\rm 5'-GGG\; CGG\; CGA\; CCT,\quad 5'-AGG\; TCG\; CCG\; CCC,$$
where A,C,G,T are the nucleotides adenine, cytosine, guanine and thymine, respectively.  Our technique relies on the 
incorporation of T's modified to include the desired thiol group~\cite{modT,LeziusRao}.  The `sticky' ends are 
filled in by a standard reaction~\cite{chemistry} using the Klenow fragment of DNA polymerase and the three 
deoxynucleoside triphosphates dATP, dGTP, and S$^4$-dTTP (see Fig.~\ref{fig1}-A).  Because of the preponderance of 
modified dTTPs (and absence of dCTP) in solution, we can incorporate a significant number of modified T's at both 
ends of each DNA molecule~\cite{mismatch}.  To prevent the Klenow fragment from excising T's that are not 
Watson-Crick matched to the template, we use a mutated form of the Klenow fragment which lacks the 3'$\rightarrow$5' 
proof-reading activity~\cite{Klenow}.

We tested the incorporation of the nucleotides into the DNA ends by a ligation assay~\cite{ligation}.  Unmodified 
$\lambda$-DNA is readily ligated by T4 DNA ligase to form multimers.  In the modified $\lambda$-DNA, however, the 
sticky ends -- now filled in by the incorporated bases -- are blunt, and multimer formation is strongly suppressed.  
The reaction products were analyzed by pulsed-field gel electrophoresis.  As shown in Fig.~\ref{fig1}-B, unmodified 
$\lambda$-DNA (``natural-$\lambda$'') was efficiently ligated (lane 3).  To control for the possibility that 
unincorporated S$^4$-dTTP inhibited ligation, $\lambda$-DNA monomers were ligated in the presence of all 3 dNTP's 
minus the polymerase (``control-$\lambda$''), and the ligation was also efficient (lane 6).  The majority of the 
thiol-modified $\lambda$-DNA (``HS-$\lambda$''), however, remained as monomers (lane 5).  This provides strong 
evidence that the protocol is effective in incorporating bases into the ends of $\lambda$-DNA.

\begin{figure}[h]   
\centerline {\epsfig{figure=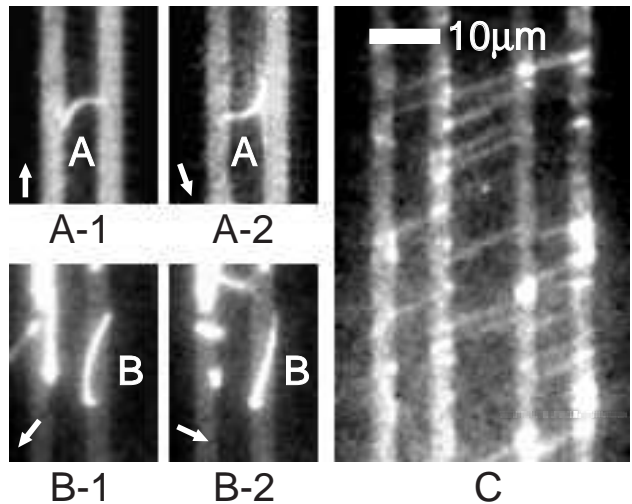,height=2.64in,width=3.25in,clip=0}} 
\caption {\label{fig2}  Images of HS-$\lambda$ DNA on quartz chips with Au electrodes. The DNA molecules were 
pre-stained with TOTO1.  Observations were through a Nikon Eclipse TE300 inverted microscope equipped with a 
60$\times$ oil-immersion lens (N.A.=1.4), and excited by a collimated Ar:Kr ion laser at 488 nm.  Fluorescence 
images were collected at 533 nm by an intensified CCD camera (RoperScientific, Princeton Instruments) and digitally 
enhanced. Au electrodes appear as vertical stripes.  Arrows represent the flow direction. Panel A-1 \& 2: Molecule 
A, anchored at both ends, and flexing with the flow.  Panel B-1 \& 2: Both ends of molecule B were attached on the 
same electrode, with the mid-segment moving freely in the buffer, showing that the attachment was specific to the 
thiol-modified ends. Panel C: Many HS-$\lambda$ DNA molecules spanning two electrodes.  The image was taken after
the Mg$^{2+}$ and NH$_4$Ac rinsing. Scale bar applies to all panels~\cite{video}.}
\end{figure}

Using standard photolithography, we constructed Au electrodes on a quartz substrate in parallel strips, 4 $\mu$m 
wide and 5 mm long, and separated by 4 or 8 $\mu$m.  The Au surfaces were rigorously cleaned~\cite{cleangold} before 
depositing the modified DNA.  At several stages during these experiments, it was important to observe the molecules 
in an optical microscope.  To image the thiol-modified DNA molecules, we stained them with the fluorescent 
intercalating dye TOTO1 (Molecular Probes), and then loaded them on the chip.  After a 20-min. incubation period, 
many of the molecules were observed to be attached to the electrodes at one end.  The unattached molecules were 
carefully rinsed in $1\times$TE~\cite{TE}.  The chip was then covered with a clean coverslip, and a flow of the 
buffer solution was applied perpendicular to the electrodes.  We observed that DNA molecules anchored at one end 
were stretched by the buffer flow to bridge the space between the electrodes.  Many of these molecules subsequently 
attached to the second electrode by their free end.  After this occurs, the flow may be repeatedly reversed to 
demonstrate that the anchored DNA molecules bow out with the flow while their ends remain anchored (Panels A and B 
of Fig.~\ref{fig2}.  See video in Ref.~\cite{video}).  This is direct evidence that chemical binding between the 
ends to Au is much stronger than physical adhesion of the rest of the molecule to either quartz or Au.  (For the 
specific DNA samples used in the resistivity measurements, we carried out the dye-staining step {\em after} the 
measurements to avoid inadvertent damage from dye intercalation.)

%

\begin{figure}[h]  
\centerline {\epsfig{figure=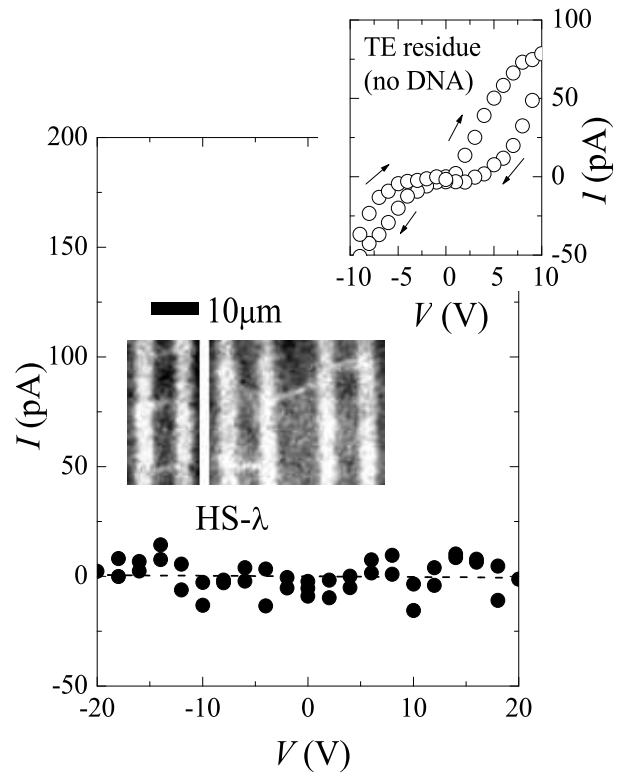,height=4.07in,width=3.14in,clip=0}} 
\caption {\label{fig3} The two-probe current vs. voltage ($I$-$V$) curve for a sample of $\lambda$-DNA bridging 2 
parallel Au electrodes separated by 4 $\mu$m (the sample comprises $\sim$1000 molecules).  The DNA was rinsed with 
NH$_4$Ac before the measurement to remove the buffer salt residue.  The dashed line is a linear fit to the data. The 
inset shows the two-probe $I$-$V$ curve for a test chip containing 1$\times$TE buffer solution (without DNA).  The 
chip was dried in vacuum but not subject to NH$_4$Ac rinsing.  The observed conductance is entirely from trace TE 
salt residue. Both measurements were done in vacuum ($<10^{-7}$ Torr) at 295 K. Open-circuit impedance between any 
two electrodes was always $\gg$10 T$\mathrm{\Omega}$.}
\end{figure}

As discussed above, a crucial step in the experiment was the removal of buffer.  In preliminary experiments, we 
repeatedly observed a finite, semiconductor-like, history-dependent conductance after loading DNA solution and 
removing the buffer solution.  The inset of Fig.~\ref{fig3} is a representative $I$-$V$ curve of the buffer salt 
residue, 1$\times$TE.  Such spurious signals are of particular concern when DNA is laid down on electrodes in a 
thick bundle, because the salts trapped between the DNA molecules can form conduction paths.  The spurious 
background vanished after we adopted the following procedure.  Chips containing bridging DNA molecules were 
carefully rinsed in 5 mM ammonium acetate (NH$_4$Ac, pH 6.6), a volatile buffer that can be completely removed in 
high vacuum.  A drawback of this rinsing is that a large fraction of the anchored DNA molecules are cut after 
rinsing and drying.  However, rinsing in 10 mM MgSO$_4$/40 mM Tris-HCl (pH 8) {\em before} the NH$_4$Ac rinsing 
introduces Mg$^{2+}$ ions which coat the quartz surface with weak positive charges~\cite{DekkerAPL}.  As the 
negatively charged DNA molecules stick to the substrate by electrostatic interaction, damage due to NH$_4$Ac
rinsing is minimized.  Panel C of Fig.~\ref{fig2} shows a typical image of anchored DNA after Mg$^{2+}$ and NH$_4$Ac 
rinsing.  [Rinsing with the MgSO$_4$ solution also led to binding of unmodified $\lambda$-DNA.  However, the yield 
of anchored molecules was much smaller.]

To perform the electrical measurements, unstained thiol-modified $\lambda$-DNA was attached to the Au electrodes as 
described above.  After the final NH$_4$Ac rinsing, the chip was dried in the dark to avoid possible photon-induced 
damage. Two-probe $I$-$V$ measurements were performed in moderately high vacuum ($<10^{-7}\mathrm{Torr}$).  

A typical room-temperature $I$-$V$ curve, measured on $\lambda$-DNA spanning electrodes 4 $\mu$m apart, is shown in 
the main panel of Fig.\ref{fig3}.  The voltage was swept between $\pm$20 V.  A linear fit to the data in 
Fig.~\ref{fig3} yields $dI/dV=(-3 \pm 9)\times 10^{-14}\mathrm{S}$.  Using a cross-section of $\sim$3 nm$^2$ per 
molecule, and the estimated number of bridging molecules ($\sim$1000), we obtain the bound on the resistivity of 
$\rho>10^6\, \mathrm{\Omega cm}$ in electric fields $E$ up to $\sim 10^4\mathrm{V/cm}$.  Measurements performed on 
several chips yielded consistent results.  No current was detected within the noise level of our 
measurement($\sim\pm$10 pA), despite sustained and deliberate efforts to improve electrical contacts between the 
base pair stack of $\lambda$-DNA and Au.

Immediately after the measurements, a buffer solution with an appropriate amount of TOTO1 dye dissolved in 
1$\times$TE was loaded on the chip.  By direct optical microscopy inspection of the post-stained DNA, we confirmed 
that there were $\sim$1000 DNA molecules bridging the electrodes, and thus the measurements did not destroy the DNA.

As a final check that the observed images are those of $\lambda$-DNA, we introduce DNase to digest the 
molecules~\cite{DNase}.  Complete deletion of all fluorescent DNA molecules was observed.  These tests leave very 
little room for doubt that a large number of {\em intact} DNA molecules were chemically bound to the electrodes 
during the electrical measurements~\cite{denature}.

The bound $\rho > 10^6\, \Omega$cm in our experiment and the large bias potential applied (20 V) is at odds with 
many recent reports of moderately high conductivity.  In some DC experiments, the DNA molecules formed bundles or 
networks between the microfabricated electrodes~\cite{ZnDNA,DNAnetwork,dyeDNA}.  As noted above, high conductance 
may arise from residual salts trapped between the DNA strands.  Contamination from other sources (C or Re) may be a 
problem as well in the experiment on proximity-induced superconductivity in $\lambda$-DNA~\cite{superconducting}. 
Microwave absorption experiments have been used to infer that $\rho\simeq  1\;\mathrm{\Omega cm}$ at 295 K in 
$\lambda$-DNA~\cite{gruner}.  The high microwave conductivity, 10$^6$ times larger than our bound, is very difficult 
to reconcile with our data.  If $\lambda$-DNA had such a high uniform conductivity, all the applied potential should 
fall across the contacts ($\sim$2-3 nm) to produce an $E$-field $\sim 10^8$ V/cm, high enough to produce a large 
tunneling current, if not breakdown of the contact barrier altogether.  This is not observed.  Possibly, the 
microwave is detecting very short dissipative regions embedded in the insulating molecule.

Two groups recently used electrostatic force microscopy to probe the electrostatic polarization of 
DNA~\cite{spanish2,Sohn}.  Our results are consistent with their conclusion that DNA is insulating.  However, the 
electrostatic force microscopy technique probes conductivity in the limit of weak bias potentials, so it does not 
rule out a transition to moderately large conductivity above a bias threshold of several volts (as reported in some 
experiments~\cite{dekker,Watanabe}).  The present experiments show that insulating behavior extends to bias 
potentials as high as 20 volts.

We are grateful to P. M. Chaikin for invaluable suggestions, and acknowledge fruitful discussions with Shirley S. 
Chan, J. Tegenfeldt, P. Silberzan, Christelle Prinz, S. Park and R. Huang.  This research is supported by a U.S. 
National Science Foundation MRSEC grant (DMR 98-09483) and by the National Inst. Health (NIH HG01506).




\begin{references}
\bibitem{modT} In the modified nucleotide 4-thiothymidine-5'-triphosphate (S$^4$-dTTP, Trilink Biotechnologies), a
sulfur atom is covalently bound to the C4 site of the thymine molecule.  Its close proximity to the hydrogen ion at
the N3 site results in a resonant state that forms the desired thiol group.
\bibitem{chemistry} The incorporation reaction was 100 $\mu$g/ml $\lambda$-DNA, 0.1 mM dATP, 0.1 mM dGTP and 0.2 mM
S$^4$-dTTP in 1$\times$ EcoPol buffer(10 mM Tris-HCl, 5 mM MgCl$_2$ and 7.5 mM dithiothreitol (DTT); pH7.5 @
25$^\circ$C; New England Biolabs), for a total volume of 100 $\mu$l. Five units of Klenow fragment
(3'$\rightarrow$5' exo$^-$) were added. The mixture was incubated at 37 $^\circ$C for 30 min. To stop the reaction,
1$\mu$l of 0.5 M EDTA was added, and the mixture heated at 75$^\circ$C for 20 min.  To exchange the buffer, the
product was dialyzed against 30 ml 0.1$\times$TE at room temperature for 45 min across a 50 nm Millipore membrane.
\bibitem{mismatch} One 5' overhang of the $\lambda$-DNA ends with an A, which is Watson-Crick matched to S$^4$-T. 
The other overhang ends with G's, which are matched to C's.  While sulfur-modified dCTP is available, the sulphur
is remote from H atoms and the resonant thiol group can not form.  To maximize the probability of modified T's at
both ends, we excluded dCTP altogether from the reaction, so that S$^4$-T:G pairing could occur in place of C:G. 
The mutated Klenow fragment used here is incapable of detecting the mis-matches.
\bibitem{Klenow} The Klenow fragment is a large fragment of DNA polymerase I lacking the error-correcting
5'$\rightarrow$3' exonuclease activity (this function damages the $\lambda$-DNA `sticky' ends).  Here, we employed
a mutated form of the Klenow fragment (3'$\rightarrow$5' exo$^-$, New England Biolabs) in which the proof-reading
3'$\rightarrow$5' exonuclease activity is deleted.  S$^4$-dTTP binds more weakly to A than unmodified
dTTP~\cite{LeziusRao}, and hence the proof-reading activity could possibly remove incorporated S$^4$-dTTP.
\bibitem{ligation} The reaction was carried out in 1$\times$ ligation buffer (50 mM Tris-HCl, 10 mM MgCl$_2$, 10 mM
DTT, 1 mM ATP and 25 $\mu$g/ml BSA; pH 7.5@25 $^\circ$C) at a $\lambda$-DNA concentration of 50 $\mu$g/ml. 400
units of T4 DNA ligase (New England Biolabs) was added to the reaction volume of 20 $\mu$l, and incubated at room
temperature for 5 h.
\bibitem{cleangold} To clean the Au surface, freshly prepared chips were soaked in fuming HNO$_3$ for 30 min, and
further cleaned in a mixture of 18 M$\mathrm{\Omega cm}$ deionized water, ammonium hydroxide and hydrogen peroxide
(5:1:1 by volume, 70 $^\circ$C) for 20 min. The chips were thoroughly rinsed in deionized and distilled water (DD
H$_2$O), and kept in Ar-flushed DD H$_2$O for no more than 30 min before loading DNA samples. The Au surface was
never allowed to dry during the process.
\end{references}
\end{document}